\documentstyle[twocolumn,pra,aps,epsf]{revtex}
\begin{document}

\draft
\title{Local-field corrections to the decay rate of excited\\ 
molecules in absorbing cavities: the Onsager model}
\author{M. S. Toma\v s}
\address{Rudjer Bo\v skovi\' c Institute, P. O. B. 180, 10002 Zagreb,
Croatia}
\date{September 13, 2000}
\maketitle
\begin{abstract}
The decay rate and the classical radiation power of an excited molecule 
(atom) located in the center of a dispersive and absorbing dielectric 
sphere taken as a simple model of a cavity are calculated adopting 
the Onsager model for the local field. The local-field correction factor 
to the external (radiation and absorption) power loss of the molecule 
is found to be $|3\varepsilon(\omega)/[3\varepsilon(\omega)+1]|^2$,
with $\varepsilon(\omega)$ being the dielectric function of the sphere.
However, local-field corrections to the total decay rate (power loss) 
of the molecule are found to be much more complex, including those to 
the decay rate in the infinite cavity medium, as derived very recently by 
Scheel {\it et al}. [Rev. A {\bf 60}, 4094 (1999)], and similiar 
corrections to the cavity-induced decay rate. The results obtained can 
be cast into model-independent forms. This suggests the general results 
for the local-field corrections to the decay rate and to the external 
power loss of a molecule in an absorbing cavity valid for molecule 
positions away from the cavity walls. 

\end{abstract}

\section{Introduction}

Within macroscopic electrodynamics, the decay rate $\Gamma$ of an 
excited molecule (atom) at a position ${\bf r}_0$ in an absorbing cavity 
is given by
\begin{equation}
\Gamma=\frac{2\omega^2}{\hbar c^2}{\bf p}^*_{fi}
\cdot{\rm Im}\tensor{\bf G}({\bf r}_0,{\bf r}_0;\omega)\cdot{\bf p}_{fi},
\label{gamagen}
\end{equation}
with $\tensor{\bf G}({\bf r},{\bf r}_0;\omega)$ being the (classical)
dyadic Green function for the system, ${\bf p}_{fi}$ the relevant 
dipole transition matrix element, and $\omega$ the transition frequency.
This result is most simply obtained using the classical theory of the 
molecular (radiative) decay \cite{ChFo} in conjuction with the 
correspondence principle and extending it straightforwardly to 
absorbing systems \cite{tomas1}. Within QED, however, it is derived 
by employing  the quantized form of the macroscopic field in absorbing 
systems obtained rather recently \cite{welsch,matloob} and 
using the Fermi golden rule \cite{barnett,barnett1} or solving 
Heisenberg's equations of motion for the molecule and the field in 
the Markov approximation \cite{scheel2,dung}.

Splitting the Green function into the translationally 
invariant part $\tensor{\bf G}^0({\bf r}-{\bf r}_0;\omega)$ and the 
scattering part $\tensor{\bf G}^{\rm sc}({\bf r},{\bf r}_0;\omega)$, 
the decay rate can be generally expressed as
\begin{equation}
\Gamma=\Gamma^0+\Gamma^{\rm sc},
\label{gama}
\end{equation}
where $\Gamma^0$ is the decay rate as would be in the infinite cavity 
medium and $\Gamma^{sc}$ is the corresponding cavity-induced decay 
rate. Upon an appropriate regularization of
$\tensor{\bf G}^0({\bf r}-{\bf r}_0;\omega)$, the rate $\Gamma^0$ is 
given by \cite{barnett1,tomas1}
\begin{equation}
\Gamma^0=\Gamma_{\rm free}\left[\frac{3}{2}\frac{\varepsilon''(\omega)}
{|\varepsilon(\omega)|^2}\left(\frac{c}{\omega R_m}\right)^3
+\eta({\omega})\right],
\label{gama0}
\end{equation}
where
\begin{equation}
\varepsilon(\omega)=[\eta({\omega})+i\kappa(\omega)]^2
\end{equation}
is the dielectric function of the cavity medium, $R_m$ an effective 
molecule-medium distance, and
\begin{equation}
\Gamma_{\rm free}=\frac{4\omega^3|{\bf p}_{fi}|^2}{3\hbar c^3}
\label{gamaf}
\end{equation}
is the free-space spontaneous emission (SE) rate. The first 
contribution to $\Gamma^0$ in Eq. (\ref{gama0}) is identified as the 
nonradiative decay rate $\Gamma^0_{\rm nr}$ due to the near-field 
mediated transfer of the molecular energy to the
surrounding medium, whereas the second one is the familiar decay rate 
$\Gamma^0_{\rm rad}$ due to the radiation losses of 
the molecule in the medium, i.e., the SE rate $\Gamma^0_{\rm SE}$ 
\cite{barnett,lee,yuze}. 

For an optically dense cavity medium, the above result has to be
improved by accounting for the difference between the macroscopic
field used in its derivation and the actual (local) field with which the 
molecule interacts. Restricting ourselves to low-density cavity media, 
in our previous consideration of the molecular decay \cite{tomas1} as well 
as of the spontaneous emission spectrum \cite{tomas2} in an absorbing
planar cavity, we ignored this difference, implicitly assuming, 
however, that the decay rate corrected for the effect of the local 
field in an isotropic cavity was of the form
\begin{equation}
\tilde{\Gamma}_{\rm loc}={\cal L}[\Gamma^0+\Gamma^{sc}],
\label{gamaL}
\end{equation}
where ${\cal L}$ is an appropriate generalization of the local-field 
correction factor in the lossless case. Thus, as suggested by 
Barnett {\it et al}. for the decay rate $\tilde{\Gamma}^0_{\rm loc}$ in 
the infinite cavity  \cite{barnett,barnett1}, one would have
\begin{equation}
{\cal L}_{\rm Lor}(\varepsilon)=
\left|\frac{\varepsilon(\omega)+2}{3}\right|^2
\label{LLor}
\end{equation}
in the Lorentz (virtul-cavity) model 
\cite{lorentz,lorenz,pantell,sipe,knoester} and 
\begin{equation}
{\cal L}_{\rm Ons}(\varepsilon)=\left|\frac{3\varepsilon(\omega)}
{2\varepsilon(\omega)+1}\right|^2
\label{LOns}
\end{equation}
in the Onsager (real-cavity) model \cite{onsager,yablo,glauber,cao}
for the local field. In either case, the effect of the local field 
would cancel when dealing with the normalized rate 
$\hat{\Gamma}=\tilde{\Gamma}_{\rm loc}/\tilde{\Gamma}^0_{\rm loc}=
\Gamma/\Gamma^0$.

The conjecture expressed by Eqs. (\ref{gamaL})-(\ref{LOns}) has 
recently been shown to be incorrect for both models for the local field, 
which we indicate in Eq. (\ref{gamaL}) by putting the tilde sign on
$\Gamma$. First, Scheel {\rm et al}. \cite{scheel1} demonstrated that, 
in the Lorentz model, a proper inclusion of the (quantum) noise 
polarization in the local field led to a more complex  
$\omega$-dependence of the decay rate $\Gamma^0_{\rm loc}$ in an 
absorbing medium than that given by the simple 
product ${\cal L}_{\rm Lor}\Gamma^0$ [see Eqs. (49) and (50) of Ref. 
\cite{scheel1}]. Their result is in full agreement with 
$\Gamma^0_{\rm loc}$ in absorbing dielectrics derived by Fleischhauer 
using a microscopic approach \cite{fleischhauer}. In a subsequent work 
Scheel {\rm et al}. \cite{scheel2} showed that the decay rate 
$\Gamma^0_{\rm loc}$ in the Onsager model for the local field
was also much more complex in the absorption case than that given by 
the product ${\cal L}_{\rm Ons}\Gamma^0$ [see Eq. (\ref{g0loc}) below]. 

Knowing the decay rate $\Gamma^0_{\rm loc}$ in an infinite 
cavity (medium), it is natural to seek for the decay 
rate $\Gamma_{\rm loc}$ in a finite cavity, i.e., in view of
Eq. (\ref{gama}), for the cavity-induced rate $\Gamma^{\rm sc}_{\rm loc}$.
Clearly, the calculation of $\Gamma^{\rm sc}_{\rm loc}$ is most
straightforwadly performed if one adopts the Onsager model for the 
local field. In this model, one assumes the molecule in the center 
of an empty spherical cavity of radius small compared with the 
transition wavelength. In view of Eq. (\ref{gamagen}), the problem then 
reduces to the calculation of the Green function for 
the system with the Onsager cavity at the source position. Owing to 
the combined symmetry of such a composite system, however, this represents 
a difficult task, which for most relevant cavity geometries demands
approximative or numerical methods \cite{girard}. Therefore, in order 
to make a (first) step towards determination of 
$\Gamma^{\rm sc}_{\rm loc}$, in this work we consider a special case 
where the molecule is located in the center of an absorbing spherical 
particle taken as a simple model of a cavity. High symmetry of this 
configuration enables one to perform a simple and exact 
calculation of $\Gamma^{\rm sc}_{\rm loc}$ in the Onsager model
as well as of the power loss $W^{\rm loc}_{\rm ext}$ of the molecule 
outside the sphere. In turn, these results (may) provide a lead to 
the corresponding solutions for a more general configuration. We note 
that this special configuration has recently been considered 
by Cao {\it et al}. \cite{cao} in their calculation of the local-field 
corrections to the SE rate in a purely dispersive medium. We partially 
adopt their approach generalizing it to an absorbing multilayered 
spherical system. It should also be noted that owing to the existence 
of high-Q resonances and, accordingly, great ability of enhancing optical 
processes \cite{chang}, dielectric microspheres are very attractive 
objects for cavity QED studies. Thus, modification of the decay rate 
and the radiation intensity of an excited molecule (atom) in, or near, 
a (lossless) microsphere has been theoretically considered in both 
the weak \cite{chew,sullivan,jhe,klimov,dung} and the strong 
\cite{klimov1,dung} molecule-field coupling limit, and experimental 
observations of modified fluorescence intensity have also been 
reported \cite{lin,barnes}.

Owing to its intuitive clarity and easy visualization, we prefer a
classical discourse in this work. Therefore, in Sec. II we recall 
the classical approach to the problem of the decaying molecule
and rederive Eq. (\ref{gama0}) raising, through a plausible argument, 
the question on the completeness of this result for absorbing media. 
In Sec. III we first obtain formal results for the decay rate and 
the power loss in the external layer of an excited molecule in the 
center of an absorbing multilayered spherical system. We then apply 
these general results to a medium with the Onsager cavity and 
provide an alternative derivation of $\Gamma^0_{\rm loc}$ 
to that given in Ref. \cite{scheel2}. Subsequently, by 
considering an absorbing dielectric sphere with and without the 
Onsager cavity, we determine $\Gamma^{\rm sc}_{\rm loc}$ and 
$W^{\rm loc}_{\rm ext}$ in terms of the corresponding quantities for 
the bare sphere. In Sec IV we briefly illustrate the effect 
of the local field in this model on $\Gamma^{\rm sc}_{\rm loc}$ and 
the total rate $\Gamma_{\rm loc}$. The main points of this work are 
summarized in Sec. V and the necessary mathematical background is 
given in Appendices A and B.

\section{Preliminaries}
In the classical approach, an excited molecule is represented by a point
dipole ${\bf p}\exp(-i\omega t)$ at the molecular position ${\bf r}_0$
oscillating with the frequency of the transition $\omega$. The molecular 
decay rate $\Gamma$ is then related through $\Gamma=W/\hbar\omega$ 
to the power
\begin{equation}
W =\frac{\omega}{2}{\rm Im}\;{\bf p}^*\cdot
{\bf E}({\bf r}_0,{\bf r}_0;\omega)
\label{Wgen}
\end{equation}
lost by the dipole in supporting its own field. Equation (\ref{gamagen})
is then obtained introducing the Green function of the system through 
\begin{equation}
{\bf E}({\bf r},{\bf r}_0;\omega)=\frac{\omega^2}{c^2}
\tensor{\bf G}({\bf r},{\bf r}_0;\omega)\cdot{\bf p},
\label{EG}
\end{equation}
noting that $\tensor{\bf G}({\bf r}_0,{\bf r}_0;\omega)$
is the diagonal dyadic and using the correspondence principle to let
${\bf p}\rightarrow 2{\bf p}_{fi}$ \cite{corrp}.

The dipole field (Green function) in a cavity can always be written
as
\begin{equation}
{\bf E}({\bf r},{\bf r}_0;\omega)={\bf E}^0({\bf R};\omega)
+{\bf E}^{sc}({\bf r},{\bf r}_0;\omega),
\label{E0Esc}
\end{equation}
where \cite{tomas1}
\begin{eqnarray}
{\bf E}^0({\bf R};\omega)&=&
\frac{1}{\varepsilon}\left[\frac{3\hat{\bf R}\hat{\bf R}-
\tensor{\bf I}}{R^3}(1-ikR)\right.\nonumber\\
&-&\left.\frac{4\pi}{3}\tensor{\bf I}\delta({\bf R})+
k^2\frac{\tensor{\bf I}-\hat{\bf R}\hat{\bf R}}{R}\right]
\cdot{\bf p}e^{ikR},
\label{E0}
\end{eqnarray}
with $\tensor{\bf I}$ being the unit dyadic, 
${\bf R}={\bf r}-{\bf r}_0$, $\hat{\bf R}={\bf R}/R$, and
\begin{equation}
k(\omega)\equiv\sqrt{\varepsilon(\omega)}\frac{\omega}{c}=
k'(\omega)+ik''(\omega),
\label{k}
\end{equation}
is the field of the dipole as would be in the infinite cavity (medium) 
and ${\bf E}^{sc}({\bf r},{\bf r}_0;\omega)$ is the component of the
dipole field scattered from the cavity walls. Owing to the 
singular longitudinal (near-field) component of 
${\bf E}^0({\bf R};\omega)$
\begin{equation}
{\bf E}^0_\parallel({\bf R};\omega)=
\frac{1}{\varepsilon}\left[\frac{3\hat{\bf R}\hat{\bf R}-
\tensor{\bf I}}{R^3}-\frac{4\pi}{3}\tensor{\bf I}\delta({\bf R})\right]
\cdot{\bf p},
\label{E0par}
\end{equation}
it is the calculation of $\Gamma^0$ 
that represents a difficult step in determining the decay rate in 
absorbing systems. 

To remove the singularity from 
${\bf E}^0_\parallel({\bf R};\omega)$ in the spirit of the 
macroscopic-field approach, one may average this component of the 
dipole field over an appropriately chosen spherical volume 
$V_m=(4\pi/3)R^3_m$ around the molecule \cite{barnett1} or regularize 
it by letting $\delta({\bf R})\rightarrow 1/V_m$ \cite{tomas1,vries}. 
In either case, one finds that
\begin{equation}
\left.{\bf E}^0_\parallel({\bf R};\omega)\right|_{R\rightarrow 0}=
-\frac{1}{\varepsilon}\frac{4\pi}{3V_m}{\bf p}.
\label{limE0par}
\end{equation}
Since for the dipole transverse field 
\[{\bf E}^0_\perp({\bf R};\omega)=
{\bf E}^0({\bf R};\omega)-{\bf E}^0_\parallel({\bf R};\omega)\] 
we find
\begin{equation}
{\bf E}^0_\perp({\bf R};\omega)=ik\frac{\omega^2}{c^2}\frac{2}{3}{\bf p}
\label{E0tr}
\end{equation}
as ${\bf R}\rightarrow 0$, this leads to the total classical dipole 
power loss 
\begin{equation}
W^0=\frac{\omega^4|{\bf p}|^2}{3c^3}
\left[\frac{3}{2}\frac{\varepsilon''(\omega)}
{|\varepsilon(\omega)|^2}\left(\frac{c}{\omega R_m}\right)^3
+\eta({\omega})\right]
\label{W01}
\end{equation}
and, accordingly, to the decay rate $\Gamma^0$ given by Eq. (\ref{gama0}).

That the above result oversimplifies the frequency dependence
of the molecular decay rate in absorbing media becomes clear
if one tries to calculate $\Gamma^0$ using the general relation
\cite{welsch}
\begin{eqnarray}
{\rm Im}&&G_{ij}({\bf r},{\bf r}_0;\omega)=\nonumber\\
&&\frac{1}{4\pi}\int d^3{\bf s}\frac{\omega^2}{c^2}
\varepsilon''({\bf s},\omega)G_{il}({\bf r},{\bf s};\omega)
G^*_{jl}({\bf r}_0,{\bf s};\omega)
\nonumber
\end{eqnarray}
and applying it to a homogeneous [$\varepsilon({\bf r},\omega)=
\varepsilon(\omega)$] medium to determine
${\rm Im}G^0_{ij}({\bf r}_0,{\bf r}_0;\omega)$.
In view of Eqs. (\ref{Wgen}) and (\ref{EG}), in a less abstract 
language this is equivalent to using the Poynting's theorem to obtain 
$W^0$. Thus, by calculating the dipole energy flow $W^0_f$ through a 
spherical surface around the dipole and the energy $W^0_a$ absorbed 
per second in the enclosed volume, we find (see Appendix A)
\begin{eqnarray}
W^0&&=\frac{\omega^4|{\bf p}|^2}{3c^3}\times\nonumber\\
&&\left[\frac{\varepsilon''}{|\varepsilon|^2}
\left|(1-ikR_c)e^{ikR_c}\right|^2(\frac{c}{\omega R_c})^3
+\eta e^{-2k''R_c}\right],
\label{Wpoy}
\end{eqnarray}
where $R_c$ is (formally) the lower limit of the radial integration 
in $W^0_a$. Owing to their characteristic dependence on the dielectric 
function of the medium, we refer to two terms in this equation
as the absorption ($\sim\varepsilon''$) and the radiation ($\sim\eta$) 
contribution to $W^0$, respectively. Since this notation may associate 
to $W^0_a$  and $W^0_f$ as the respective origins of these contributions, 
we stress that both $W^0_a$ and $W^0_f$ are needed to obtain each of 
them, as is clear from the derivation in Appendix A.

To obtain the dipole power loss from Eq. (\ref{Wpoy}),
the $R_c\rightarrow 0$ limit should eventually be taken. However, 
consider $R_c$ as a small ($R_c\ll \lambda$) but finite cutoff for the
moment. In this case, expanding $W^0$ in powers of $\omega R_c/c$, 
we find [to ${\cal O}(\frac{\omega R_c}{c})$]
\begin{eqnarray}
\tilde{W}^0&=&\frac{\omega^4|{\bf p}|^2}{3c^3}
\left\{\frac{\varepsilon''(\omega)}
{|\varepsilon(\omega)|^2}
\left[(\frac{c}{\omega R_c})^3+\varepsilon'(\omega)
\frac{c}{\omega R_c}\right.\right.\nonumber\\
&&\left.\left.-\frac{2}
{3}[\eta(\omega)\varepsilon''(\omega)+
\kappa(\omega)\varepsilon'(\omega)]\right]+\eta(\omega)\right\}.
\label{W02}
\end{eqnarray}
In addition to a near-field term, which very much resembles the 
corresponding term in Eq. (\ref{W01}), now we have two new absorption 
terms. The most striking is the appearance of another $R_c$-free 
term which, therefore, persists even in the $R_c\rightarrow 0$ limit. 
This implies that a corresponding term must appear in any calculation 
of $W^0$, which indicates that Eq. (\ref{W01}) is, in this respect, 
incomplete.

Taking $R_c$ in Eq. (\ref{Wpoy}) as a cutoff is equivalent to setting 
${\bf E}^0({\bf R};\omega)=0$ for $R<R_c$. Since this field does not 
obey Maxwell's equations, it is clearly incorrect to regard
$\tilde{W}^0$ as the dipole power loss, which we have emphasized by 
using the tilde. On the other hand, extending the macroscopic field 
down to intermolecular distances ($R_c\rightarrow 0$) is not justified 
as the actual field acting on the molecule may largely differ from it. 
These ambiguities concerning $W^0$ are naturally resolved within an 
exact macroscopic-field approach in the following section, where we 
adopt the Onsager (real cavity) model for the local field and
therefore assume the molecule in the center of an empty spherical 
cavity with the radius $R_c$ small compared with the transition 
wavelength $\lambda$. Since, in this case, the longitudinal component
${\bf E}^0_\parallel({\bf R};\omega)$ of the dipole field does not
contribute to the molecular power loss, no
singularity appears in the theory.

\section{Local-field corrections}

Consider an excited molecule (dipole) in the center of an 
$N$-layered spherical system, as depicted in Fig. 1. 
In this case, the power loss $W^{\rm sc}$ of the molecule  may be 
written as
\begin{equation}
W^{\rm sc}=\frac{\omega}{2}{\rm Im}\;
\left.{\bf p}^*\cdot{\bf E}^{\rm sc}_1({\bf r};\omega)\right|
_{r\rightarrow 0},
\label{gscgen}
\end{equation}
where ${\bf E}^{\rm sc}_1({\bf r};\omega)=
{\bf E}^{\rm sc}_1({\bf r},{\bf r}_0;\omega)|_{r_0\rightarrow 0}$ is 
the scattered part of the dipole field in the central region. 
The calculation of the dipole field in this configuration
is outlined in Appendix B, assuming, for simplicity, that 
${\bf p}=p\hat{\bf z}$. Using Eq. (\ref{E1sc0}), we find from 
Eq. (\ref{gscgen}) that ($k_0=\omega/c$)
\begin{equation}
W^{\rm sc}=W_{\rm free}{\rm Re}\sqrt{\varepsilon_1}C^N_1,\;\;\;
W_{\rm free}=\frac{ck_0^4|{\bf p}|^2}{3},
\label{Wsc}
\end{equation}
where $C^N_1$ is the corresponding reflection coefficient. 
In the case of the empty central region [$W^0=W_{\rm free}$], we 
therefore have for the normalized total decay rate 
$\hat{\Gamma}=W/W_{\rm free}$ of the molecule
\begin{equation}
\hat{\Gamma}=1+{\rm Re}\left[C^N_1\right]_{\varepsilon_1=1}.
\label{ggen}
\end{equation}
Identifying the central sphere with the Onsager cavity, this 
general result provides a direct way for inclusion of the local-field 
corrections to the decay rate in spherical multilayered systems.
\begin{figure}[htb]
\begin{center}
\leavevmode
\hbox{%
\epsfxsize=8.6cm
\epsffile{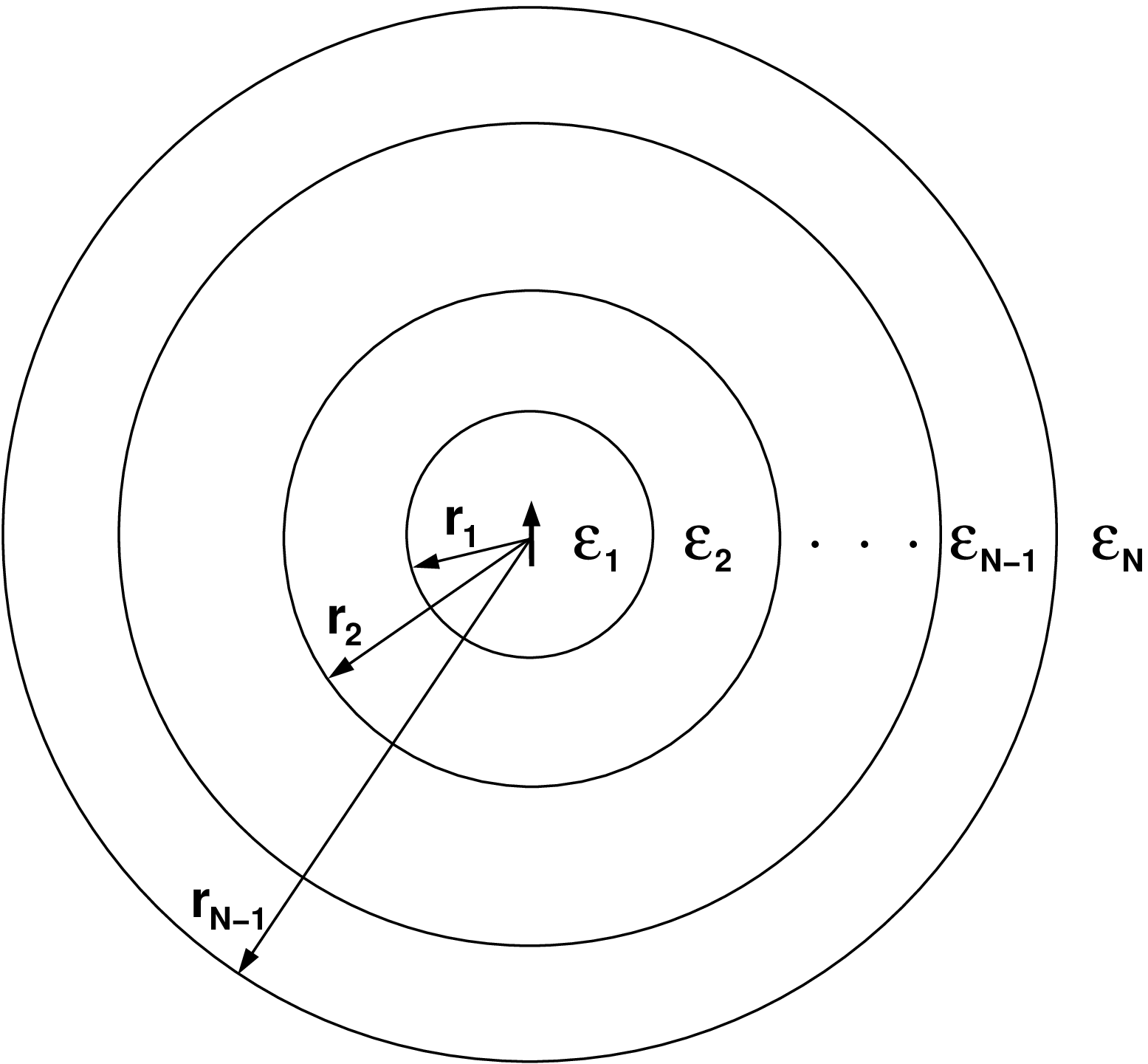}}
\end{center}
\caption{System considered in this paper. All layers are assumed 
absorbing and desribed by the complex dielectric function 
$\varepsilon_i(\omega)$.}
\end{figure}

Of obvious interest is also the radiation power $W_N^{\rm rad}$ of 
the molecule or, generally, the total power loss $W_N$ in the outer 
region of the multilayer. By comparing Eqs. (\ref{E0sph}) and (\ref{El}), 
we see that the dipole field ${\bf E}_N({\bf r};\omega)$ in this region 
is the same as the field produced in the infinite medium ($N$) of the 
dipole ($C^N_N\equiv C^N_{N+}$):
\begin{equation}
{\bf p}_N=\frac{\varepsilon_1}{\varepsilon_N}C^N_N{\bf p}.
\label{pN}
\end{equation}
Accordingly, provided that we let ${\bf p}\rightarrow {\bf p}_N$, 
$\varepsilon\rightarrow\varepsilon_N$, and $R_c\rightarrow r_{N-1}$, 
we can adopt all results concerning the dipole power loss in an infinite 
medium derived in Appendix A. For example, with these replacements, 
the angular distribution of radiation $dW^{\rm rad}_N/d\Omega$
is obtained by keeping only the radiation field ($\sim 1/r^2$)
contribution to Eq. (\ref{Prad}) 
\begin{equation}
\frac{dW^{\rm rad}_N}{d\Omega}=\eta_N\frac{ck_0^4|{\bf p}|^2}{8\pi}
\left|\frac{\varepsilon_1}{\varepsilon_N}C^N_N\right|^2
e^{-2k''_Nr}\sin^2\vartheta,
\end{equation}
the radiation power $W^{\rm rad}_N$ is given by the last term in
Eq. (\ref{W0flow}) and the total power loss $W_N$ by Eq. (\ref{Wpoy}).
Regarding the central sphere as the Onsager cavity, we see that 
the local-field correction factor to $W_N$ (and $W^{\rm rad}_N$) 
is given by $|C^N_N/\varepsilon_N|^2$ for $\varepsilon_1=1$ and 
in the limit $k_0a_1=k_0R_c\ll 1$.

\subsection{Infinite medium}

To calculate local-field corrections to the decay rate in an infinite 
medium with the dielectric function $\varepsilon(\omega)$, we consider 
a two-layered system consisting of this medium with the Onsager 
cavity cut around the origin. Then, as was done by Scheel {\it et al}. 
\cite{scheel2}, $\hat{\Gamma}^0_{\rm loc}$ is straightforwardly obtained
using Eq. (\ref{ggen}) and expanding the reflection coefficient 
$C^2_1(1,\varepsilon;R_c)$ given by Eq. (\ref{C12}) in powers of 
$\rho_1=k_0R_c\ll 1$. It is very instructive to rederive 
this result by applying the Poynting's theorem to a spherical surface 
around the molecule enclosing the Onsager cavity and thus determine 
the power loss $W^0_{\rm loc}$ of the molecule. We note that we have 
performed such a calculation in Appendix A to obtain 
Eq. (\ref{Wpoy}). Thus, $W^0_{\rm loc}$ is given by this 
equation provided that we let ${\bf p}\rightarrow{\bf p}_{\rm eff}$, 
where [cf. Eq. (\ref{pN})] 
\[
{\bf p}_{\rm eff}=\frac{1}{\varepsilon}C^2_2(1,\varepsilon;R_c){\bf p},
\]
and regard $R_c$ as the Onsager cavity radius.

Expanding the coefficient $C^2_2(1,\varepsilon;R_c)$ 
given by Eq. (\ref{C22}) for small $\rho_1=k_0R_c$, we find
\begin{eqnarray}
{\bf p}_{\rm eff}&=&\frac{3\varepsilon}{2\varepsilon+1}
\left[1-\frac{10\varepsilon^2-9\varepsilon-1}{10(2\varepsilon+1)}
(k_0R_c)^2\right.\nonumber\\
&&\left.-i\frac{2}{3}\frac{\varepsilon^\frac{3}{2}(\varepsilon-1)}
{2\varepsilon+1}(k_0R_c)^3
+{\cal O}[(k_0R_c)^4]\right]{\bf p}.
\label{peff}
\end{eqnarray}
Owing to the $(k_0R_c)^{-3}$ factor in the first term on the rhs
of Eq. (\ref{Wpoy}), in this term we must use
\begin{eqnarray}
&&\left|(1-i\sqrt{\varepsilon}k_0R_c)e^{i\sqrt{\varepsilon}k_0R_c}
{\bf p}_{\rm eff}\right|^2=\nonumber\\
&&\left|\frac{3\varepsilon}{2\varepsilon+1}\right|^2
\left[1+\frac{1}{5}{\rm Re}\frac{14\varepsilon+1}{2\varepsilon+1}
(k_0R_c)^2\right.\nonumber\\
&&\left.-2{\rm Im}\frac{\varepsilon^\frac{3}{2}}
{2\varepsilon+1}(k_0R_c)^3
+{\cal O}[(k_0R_c)^4]\right]|{\bf p}|^2.
\end{eqnarray}
Since the second term on the rhs of Eq. (\ref{Wpoy}) is a 
well-behaved function of $k_0R_c$, in this term it is sufficient 
to let ${\bf p}_{\rm eff}=3\varepsilon/(2\varepsilon+1){\bf p}$. 
In this way, for the normalized decay rate 
$\hat{\Gamma}^0_{\rm loc}=W^0_{\rm loc}/W_{\rm free}$ we obtain 
[to ${\cal O}(\frac{\omega R_c}{c})$]
\begin{eqnarray}
\hat{\Gamma}^0_{\rm loc}&=
&\left|\frac{3\varepsilon(\omega)}{2\varepsilon(\omega)+1}\right|^2
\left\{\eta(\omega)+\frac{\varepsilon''(\omega)}
{|\varepsilon(\omega)|^2}
\left[(\frac{c}{\omega R_c})^3\right.\right.\nonumber\\
&+&\frac{28|\varepsilon(\omega)|^2+16\varepsilon'(\omega)+1}
{5|2\varepsilon(\omega)+1|^2}(\frac{c}{\omega R_c})\nonumber\\
&-&\left.\left.2\frac{2\kappa(\omega)|\varepsilon(\omega)|^2+
\kappa(\omega)\varepsilon'(\omega)+\eta(\omega)\varepsilon''(\omega)}
{|2\varepsilon(\omega)+1|^2}\right]\right\}.
\label{g0loc}
\end{eqnarray}

The above result coincides with that of Scheel {\it et al}. 
\cite{scheel1}. 
This time, however, the origin of various contributions to
$\hat{\Gamma}^0_{\rm loc}$ along with their separate local-field 
corrections can be clearly identified. As seen from comparison
with Eq. (\ref{W01}), while the near-field and the radiation-field 
terms get multiplied by ${\cal L}_{\rm Ons}$ as expected \cite{barnett1}, 
it is the appearance of the additional absorption terms already signaled 
in Eq. (\ref{W02}) that represents essentially new corrections to 
$\hat{\Gamma}^0$ in absorbing media. Of these two terms, the most 
interesting is the (usually) negative $R_c$-free contribution. This term
effectively adds to the radiation-field contribution ($\eta$) to the 
decay rate in absorbing media and tends to diminish the overall rate.

\subsection{Cavity}
Having determined $\Gamma^0_{\rm loc}$, the next step is to consider
the decay rate $\Gamma_{\rm loc}$ in the general case when the molecule 
is embedded in an inhomogeneous system, i.e., in a cavity. In order to 
determine $\Gamma_{\rm loc}$, we consider the decaying molecule in the 
center of a dielectric sphere of radius $R$ and the dielectric 
function $\varepsilon(\omega)$ immersed in an external medium with 
the dielectric function $\varepsilon_{\rm ext}(\omega)$. In this case,
the relevant reflection coefficient to be inserted in Eq. (\ref{ggen}) is 
$C^3_1(1,\varepsilon,\varepsilon_{\rm ext};R_c,R)$ [Eq. (\ref{C13})] 
corresponding to the sphere with the Onsager cavity. 
Expanding this coefficient in powers of $\rho_{11}=k_0R_c$, we find that
\begin{eqnarray}
&&C^3_1(1,\varepsilon,\varepsilon_{\rm ext};R_c,R)=\nonumber\\
&&-i\frac{9\varepsilon}
{2\varepsilon+1}\left(k_0R_c\right)^{-3}-
i\frac{9\varepsilon(8\varepsilon+1)}
{5(2\varepsilon+1)^2}\left(k_0R_c\right)^{-1}\nonumber\\
&&-\frac{9\varepsilon^{\frac{5}{2}}}{(2\varepsilon+1)^2}
\frac{\beta_1-\beta_2}{\beta_1+\beta_2}-1+{\cal O}(k_0R_c),
\label{C13expan}
\end{eqnarray}
with $\beta_j$ given by Eq. (\ref{betaj}). One may recognize that
\[\frac{2\beta_1}
{\beta_1+\beta_2}=-C^2_1(\varepsilon,\varepsilon_{\rm ext};R),\]
where $C^2_1(\varepsilon,\varepsilon_{\rm ext};R)$ [Eq. (\ref{C12})]
is the reflection coefficient of the system without the Onsager cavity. 
One may also see that
\begin{eqnarray}
{\rm Re}\frac{9\varepsilon^{\frac{5}{2}}}
{(2\varepsilon+1)^2}&=&
\left|\frac{3\varepsilon}{2\varepsilon+1}\right|^2\eta\nonumber\\
&-&\frac{18\varepsilon''}{|2\varepsilon+1|^4}
[(2|\varepsilon|^2+\varepsilon')\kappa+\varepsilon''\eta]
\label{Rep}
\end{eqnarray}
is equal to the $R_c$-free  contribution to $\hat{\Gamma}^0_{\rm loc}$ 
[Eq. (\ref{g0loc})] and that analogous results hold for the real parts 
of the first two terms in Eq. (\ref{C13expan}). Therefore, 
from Eq. (\ref{ggen}) we find
\begin{equation}
\hat{\Gamma}_{\rm loc}=\hat{\Gamma}^0_{\rm loc}+
\hat{\Gamma}^{\rm sc}_{\rm loc},
\label{gloc}
\end{equation}
where
\begin{equation}
\hat{\Gamma}^{\rm sc}_{\rm loc}=
{\rm Re}\frac{9\varepsilon^{\frac{5}{2}}}
{(2\varepsilon+1)^2}C^2_1(\varepsilon,\varepsilon_{\rm ext};R)
\label{gscloc}
\end{equation}
is the normalized cavity-induced decay rate with the local-field 
corrections. 

The above result for $\hat{\Gamma}^{\rm sc}_{\rm loc}$ can be 
transformed into a form similiar to Eq. (\ref{g0loc}) for
$\hat{\Gamma}^0_{\rm loc}$. Letting 
$\sqrt{\varepsilon}\rightarrow\sqrt{\varepsilon}C^2_1
(\varepsilon,\varepsilon_{\rm ext};R)$ in Eq. (\ref{Rep}), we 
see that $\hat{\Gamma}^{\rm sc}_{\rm loc}$ is obtained from 
the $R_c$-free contribution to $\hat{\Gamma}^0_{\rm loc}$ upon 
replacements $\eta\rightarrow\hat{\Gamma}^{\rm sc}$ and 
$\kappa\rightarrow 2\hat{\Delta}^{\rm sc}$, where
\begin{equation}
\hat{\Gamma}^{\rm sc}={\rm Re}\sqrt{\varepsilon_1}
C^2_1(\varepsilon,\varepsilon_{\rm ext};R)
\label{gamasc}
\end{equation}
is the normalized cavity-induced decay rate [Eq. (\ref{Wsc})]
and  
\begin{equation}
\hat{\Delta}^{\rm sc}=\frac{1}{2}{\rm Im}\sqrt{\varepsilon_1}
C^2_1(\varepsilon,\varepsilon_{\rm ext};R)
\label{deltasc}
\end{equation}
is the normalized classical cavity-induced level shift \cite{shift}
of the molecule in the bare sphere. Accordingly,
from Eq. (\ref{g0loc}) we finally have
\begin{eqnarray}
\hat{\Gamma}^{\rm sc}_{\rm loc}=&&
\left|\frac{3\varepsilon(\omega)}{2\varepsilon(\omega)+1}\right|^2
\left\{\hat{\Gamma}^{\rm sc}-2\frac{\varepsilon''(\omega)}
{|\varepsilon(\omega)|^2}\times\right.\nonumber\\
&&\left.\frac{2\left[2|\varepsilon(\omega)|^2
+\varepsilon'(\omega)\right]\hat{\Delta}^{\rm sc}+
\varepsilon''(\omega)\hat{\Gamma}^{\rm sc}}
{|2\varepsilon(\omega)+1|^2}\right\}.
\label{ggscloc}
\end{eqnarray}
We note that this is not an unexpected result once we 
have learned the correct form of the decay rate 
$\hat{\Gamma}^0_{\rm loc}$ in the infinite medium. In its derivation 
it is implicitly assumed that the radius of the sphere, i.e., the 
molecule-mirror distance, is much larger than the transition 
wavelength ($k_0R\gg 1$). Under these circumstances, the 
molecule-mirror interaction goes through the radiation-field 
component of the scattered field and therefore only this field 
component determines the cavity-induced rate. Thus, the rate
${\cal L}_{\rm Ons}\hat{\Gamma}^{\rm sc}$, as would be obtained 
by letting 
${\bf p}\rightarrow 3\varepsilon/(2\varepsilon+1){\bf p}$ 
in Eq. (\ref{Wsc}), is corrected for an absorption contribution
in the same way as is the radiation-field contribution to the rate 
$\hat{\Gamma}^0_{\rm loc}$ in the infinite cavity.

\subsection{External region}
To find the local-field corrections to the molecular losses in the 
external region, we consider the field 
${\bf E}^{\rm loc}_{\rm ext}({\bf r};\omega)$
outside the sphere with the Onsager cavity. As already noted, this 
field is equal to the field in the infinite external medium 
[Eq. (\ref{E0sph}), with $k=k_{\rm ext}$] of the dipole 
[cf. Eq. (\ref{pN})]
\begin{equation}
{\bf p}^{\rm loc}_{\rm ext}=\frac{1}{\varepsilon_{\rm ext}}
C^3_3(1,\varepsilon,\varepsilon_{\rm ext};R_c,R){\bf p},
\end{equation}
with the coefficient $C^3_3(1,\varepsilon,\varepsilon_{\rm ext};R_c,R)$ 
given by Eq. (\ref{C33}). For small $\rho_{11}=k_0R_c$, 
we find that [to ${\cal O}[(k_0R_c)^2$]]
\begin{equation}
C^3_3(1,\varepsilon,\varepsilon_{\rm ext};R_c,R)=
-\frac{i\varepsilon_{\rm ext}}{\sqrt{\varepsilon}k_0R}
\frac{3\varepsilon^2}{2\varepsilon+1}\frac{2}{\beta_1+\beta_2},
\end{equation}
with $\beta_j$ given by Eq. (\ref{betaj}). Now
\begin{equation}
\frac{i\varepsilon_{\rm ext}}{\sqrt{\varepsilon}k_0R}
\frac{2}{\beta_1+\beta_2}=-C^2_2(\varepsilon,\varepsilon_{\rm ext};R),
\end{equation}
where $C^2_2(\varepsilon,\varepsilon_{\rm ext};R)$ is the external-field
coefficient of the system without the Onsager cavity [cf. Eq. (\ref{C22})].
To the leading term in $k_0R_c$, we therefore have 
\begin{equation}
{\bf p}^{\rm loc}_{\rm ext}=\frac{3\varepsilon}{2\varepsilon+1}
\frac{\varepsilon}{\varepsilon_{\rm ext}}
C^2_2(\varepsilon,\varepsilon_{\rm ext};R){\bf p}=
\frac{3\varepsilon}{2\varepsilon+1}{\bf p}_{\rm ext},
\label{plocext}
\end{equation}
where ${\bf p}_{\rm ext}$ is the corresponding effective dipole moment 
for the bare sphere [cf. Eq. (\ref{pN})]. Accordingly, for the
external field we have
\begin{equation}
{\bf E}^{\rm loc}_{\rm ext}({\bf r};\omega)=
\frac{3\varepsilon(\omega)}{2\varepsilon(\omega)+1}
{\bf E}_{\rm ext}({\bf r};\omega),
\label{Elocext}
\end{equation}
with ${\bf E}_{\rm ext}({\bf r};\omega)$ being the external field for the 
bare sphere. Consequently, this implies ${\cal L}_{\rm Ons}$ as the proper
local-field correction factor to the external power loss 
\begin{equation}
W^{\rm loc}_{\rm ext}=\left|\frac{3\varepsilon(\omega)}
{2\varepsilon(\omega)+1}\right|^2W_{\rm ext}.
\label{Wlocext}
\end{equation}
Note that here $W_{\rm ext}$ is generally given by Eq. (\ref{Wpoy}), 
with $k=k_{\rm ext}$ and $R_c=R$. 

We end this section with a remark on a (plausible) generalization
of the obtained results. Using Eqs. (\ref{EG}) and (\ref{E1sc0}), 
the normalized cavity-induced decay rate [Eq. (\ref{gamasc})]
and frequency shift [Eq. (\ref{deltasc})] can be written in terms 
of the Green function for the bare sphere as
\begin{eqnarray}
\hat{\Gamma}^{\rm sc}&=&\frac{3}{2k_0}
\hat{\bf p}\cdot{\rm Im}\tensor{\bf G}^{\rm sc}
({\bf r}_0,{\bf r}_0;\omega)
\cdot\hat{\bf p},\nonumber\\
\hat{\Delta}^{\rm sc}&=&-\frac{3}{4k_0}
\hat{\bf p}\cdot{\rm Re}\tensor{\bf G}^{\rm sc}
({\bf r}_0,{\bf r}_0;\omega)
\cdot\hat{\bf p},
\label{gadesc} 
\end{eqnarray}
where $\hat{\bf p}$ gives the direction of the transition.
With this inserted in Eqs. (\ref{ggscloc}), any reference 
to the specific system considered in the derivation of this equation
is formally lost. Therefore, at first sight, there is no  
reason why Eqs. (\ref{ggscloc}) and (\ref{gadesc}) should not 
be taken as the general result for the molecular decay rate 
in an absorbing cavity valid for all positions of the molecule away 
from the cavity walls and for all transition dipole orientations. 
Similarly, in view of Eq. (\ref{EG}), it appears that Eq. (\ref{Elocext}) 
indicates a rather general relationship between the Green function 
elements for a system with and without the Onsager cavity:
\begin{equation}
\tensor{\bf G}_{\rm loc}({\bf r},{\bf r}_0;\omega)=
\frac{3\varepsilon(\omega)}{2\varepsilon(\omega)+1}
\tensor{\bf G}^{\rm sc}({\bf r},{\bf r}_0;\omega),
\label{GlG} 
\end{equation} 
where ${\bf r}$ is in the external layer and ${\bf r}_0$ in the cavity.
If this relation holds, ${\cal L}_{\rm Ons}(\varepsilon)$ as the 
local-field correction factor for the molecular external power loss 
in the general case would then be its immediate consequence. 
Of course, as already stressed, these conjectures cannot be proved 
without the calculation of the exact Green function 
for the system including the Onsager cavity at the source position.

\section{Discussion}
To illustrate the effect of the local field on the decay rate in 
absorbing cavities, we exploit the above simple model and consider 
the decaying molecule in the center of an absorbing dielectric
sphere surrounded by air ($\varepsilon_{\rm ext}=1$). The 
dielectric function of the sphere is modeled as
\begin{equation}
\varepsilon(\omega)=\varepsilon_b+\frac{\Omega^2}
{\omega^2_0-\omega^2-i\omega\gamma},
\label{eps}
\end{equation}
where $\varepsilon_b$ is the background (high-frequency) dielectric 
constant, $\omega_0$ and $\gamma$ are, respectively, the center 
frequency and the width of the absorption resonance and its strength 
is controlled by $\Omega^2$. In this analysis, a relatively large 
background dielectric constant $\varepsilon_b=5$ is chosen to 
strengthen the cavity effect of the sphere. 
\begin{figure}[htb]
\begin{center}
\leavevmode
\hbox{%
\epsfxsize=8.6cm
\epsffile{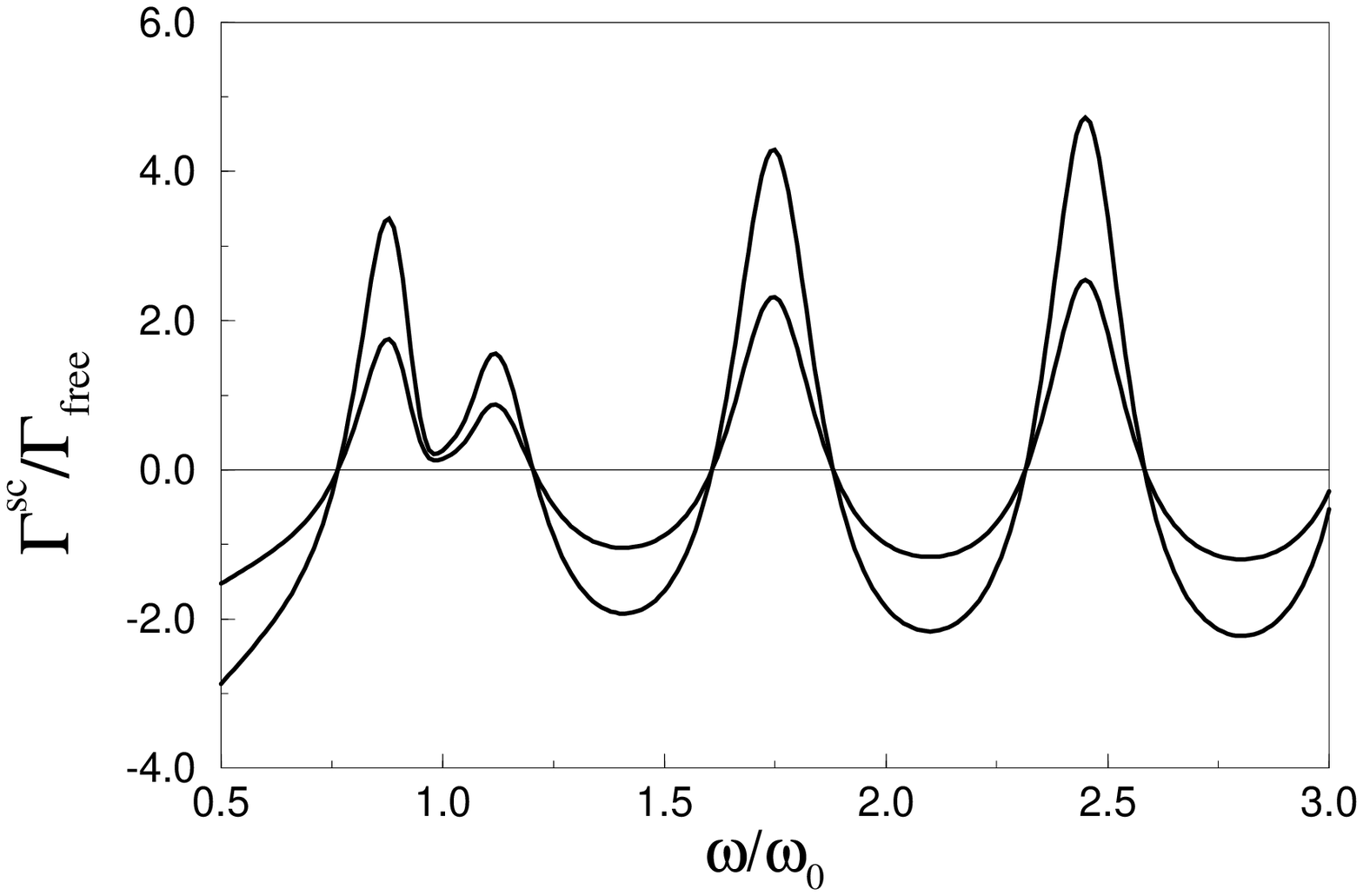}}
\end{center}
\caption{Cavity-induced decay rate with (upper curve) and without 
(lower curve) local-field corrections. The radius of the sphere is 
$R=\lambda_0/\pi$ and the material parameters are $\varepsilon_b=5$, 
$\Omega=0.5\omega_0$ and $\gamma=0.1\omega_0$.}
\end{figure}

The effects of the local field on the cavity-induced rate are 
illustrated in Fig. 2, where the rates 
$\hat{\Gamma}^{\rm sc}_{\rm loc}$ [Eq.(\ref{gscloc})] and 
$\hat{\Gamma}^{\rm sc}$ [Eq.(\ref{gamasc})] are compared for the 
system with (upper line) and without (lower line) the Onsager 
cavity, respectively. The radius ($R\omega_0/c=2$) of the sphere 
(nearly) corresponds to the first peak of 
$\hat{\Gamma}^{\rm sc}_{\rm loc}$ considered as a function 
of $\omega R/c$ in the lossless ($\Omega=0$) case and for 
$\omega=\omega_0$. For these parameters, the sphere therefore acts as 
an enhancement cavity with respect to the SE rate at $\omega=\omega_0$ 
and, since the molecule decays only radiatively, the decay rate is equal
to the SE rate. Away from the resonance, the system is lossless and, 
owing to the cavity effect of the sphere, the two (SE) rates exhibit 
the familiar oscillations with $\omega$ with the amplitudes of 
oscillations scaled by the usual local-field correction factor 
${\cal L}_{\rm Ons}(\varepsilon_b)\simeq 1.85$ 
[cf. Eq. (\ref{ggscloc})]. 
A somewhat different effect of the local field is observed in the 
region of the resonance $\omega\simeq\omega_0$. For the
medium parameters chosen, the second (absorption) term on the rhs of 
Eq. (\ref{ggscloc}) is small compared with the first one. Moreover, 
according to our calculations, $\hat{\Gamma}^{\rm sc}_{\rm loc}$ 
cannot be distinguished on this scale from 
${\cal L}_{\rm Ons}\hat{\Gamma}^{\rm sc}$ over a wide range of the 
parameters $\Omega$ and $\gamma$. The cavity-induced rate in this 
frequency region is therefore (again) predominantly determined by 
the product of ${\cal L}_{\rm Ons}$ and $\hat{\Gamma}^{\rm sc}$. Each 
of these quantities exhibits a characteristic asymmetric dispersion 
around $\omega_0$. In addition, the sphere is (nearly) in resonance 
with the medium. As a consequence, instead of a peak at 
$\omega=\omega_0$ appearing in the nonabsorbing 
($\Omega=0$) case, owing to the resonant absorption, the rates 
$\hat{\Gamma}^{\rm sc}_{\rm loc}$ and $\hat{\Gamma}^{\rm sc}$ exhibit 
asymmetric double-peak structures, which very much resemble the
"cavity-polariton" part of the SE spectrum in the bad-cavity case 
\cite{tomas2}.
\begin{figure}[htb]
\begin{center}
\leavevmode
\hbox{%
\epsfxsize=8.6cm
\epsffile{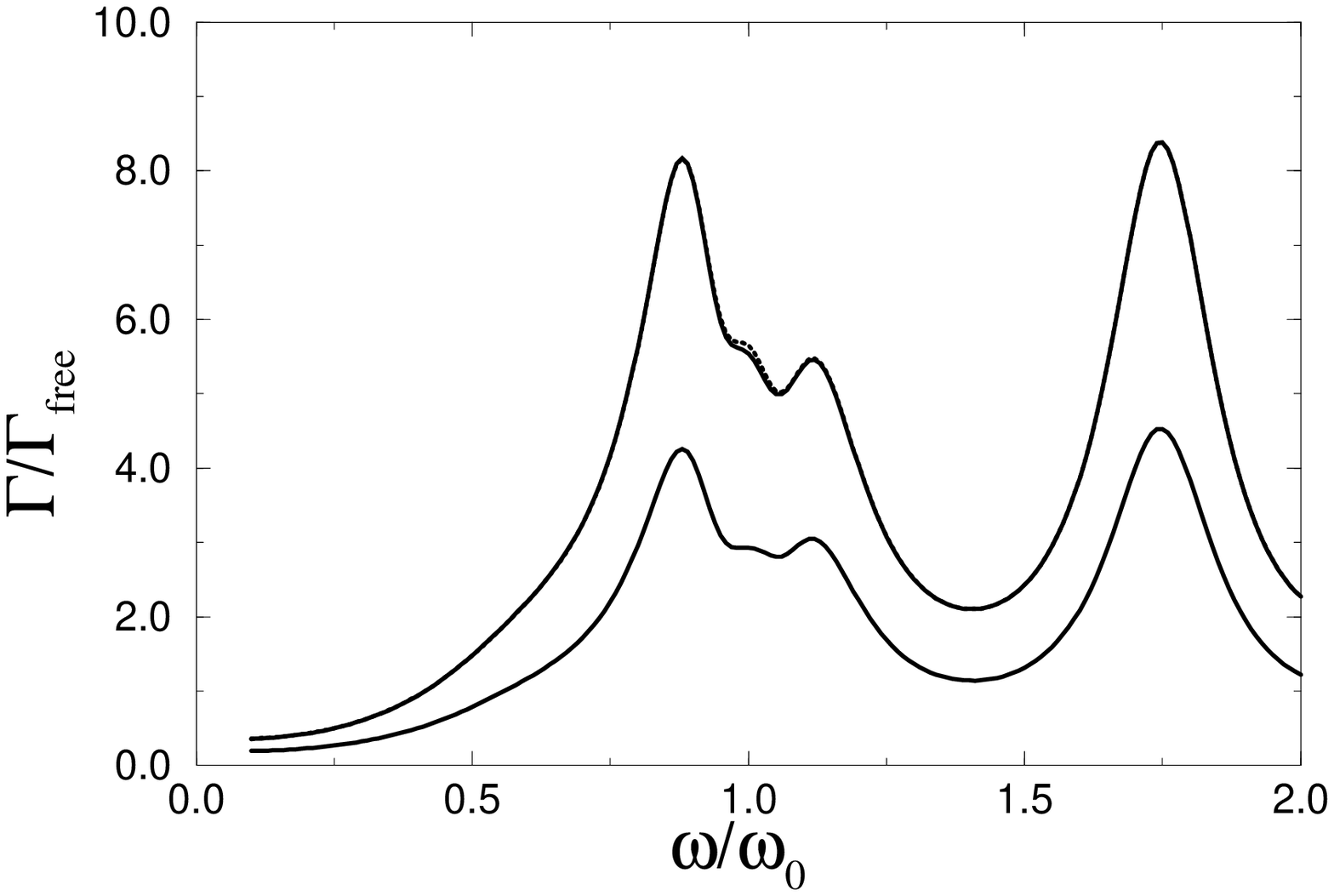}}
\end{center}
\caption{Total decay rate with (upper solid line) and without 
(lower solid line) local-field corrections for the system described 
in Fig. 2. The dotted line represents the macroscopic decay rate
with the usual local-field correction factor. 
The Onsager cavity radius is $R_c=0.1\lambda$ and $R_m=R_c$.}
\end{figure}

Figure 3 represents the total decay rates $\hat{\Gamma}_{\rm loc}$ 
[Eq. (\ref{gloc}) in conjuction with Eqs. (\ref{g0loc}) and (\ref{gscloc})] 
and $\hat{\Gamma}$ [Eq. (\ref{gama}) in conjuction with Eqs. 
(\ref{gama0}) and (\ref{gamasc})] with and without the local-field 
corrections, respectively, along with the decay rate 
$\hat{\tilde{\Gamma}}_{\rm loc}={\cal L}_{\rm Ons}\hat{\Gamma}$  
[Eqs. (\ref{gamaL}) and (\ref{LOns})] with the local-field correction 
factor assumed in our previous work \cite{tomas1,tomas2}. In plotting 
these curves, we have taken that the molecule-medium distance $R_m$ 
figuring in $\hat{\Gamma}^0$ [Eq. (\ref{gama0})] is equal to the 
Onsager cavity radius, $R_m=R_c$.
Away from the resonance, $\hat{\Gamma}_{\rm loc}$ and 
$\hat{\Gamma}$ merely reproduce the corresponding curves of Fig. 2,
with values enhanced by the off-resonance infinite-cavity decay rates
$\hat{\Gamma}^0_{\rm loc}={\cal L}_{\rm Ons}(\varepsilon_b)
\sqrt{\varepsilon_b}\simeq 4.1$ and 
$\hat{\Gamma}^0=\sqrt{\varepsilon_b}\simeq 2.2$,
respectively. As before, the situation around $\omega_0$ is different.
For this $R_c$ and the medium parameters chosen, the absorption 
contributions to $\hat{\Gamma}^0_{\rm loc}$ [Eq. (\ref{g0loc})] and 
$\hat{\Gamma}^{\rm sc}_{\rm loc}$ [Eq. (\ref{ggscloc})] are (still) 
small compared with the corresponding radiation contributions.  
Accordingly,
\begin{equation}
\hat{\Gamma}_{\rm loc}\approx{\cal L}_{\rm Ons}(\eta+
\hat{\Gamma}^{\rm sc})
\label{gloca}
\end{equation}
holds. For $R_m\approx R_c$, the same conclusion applies to the
decay rate $\hat{\Gamma}$ as well, so that 
\begin{equation}
\hat{\Gamma}\approx\eta+\hat{\Gamma}^{\rm sc}.
\label{ga}
\end{equation}
The different shapes of the curves in Figs. 3 when compared with those 
in Fig. 2 in this region are therefore caused by the (anomalous) 
dispersion of the superimposed ${\cal L}_{\rm Ons}\eta$ and $\eta$, 
respectively. This also explains why $\hat{\Gamma}_{\rm loc}$ 
(upper solid line) practically cannot be distinguished on this scale 
from the decay rate 
$\hat{\tilde{\Gamma}}_{\rm loc}={\cal L}_{\rm Ons}\hat{\Gamma}$ (dotted 
line), although the corresponding absorption contributions to these 
rates are different [cf. Eqs. (\ref{g0loc}) and (\ref{gama0})]. 
We note that Eqs. (\ref{gloca}) and (\ref{ga}) hold for a range 
of the Onsager cavity radia and/or medium parameters. In these 
cases, our previous conjecture concerning the local-field corrections 
to the decay rate, as expressed by Eq. (\ref{gamaL}), remains 
(approximately) valid.
\begin{figure}[htb]
\begin{center}
\leavevmode
\hbox{%
\epsfxsize=8.6cm
\epsffile{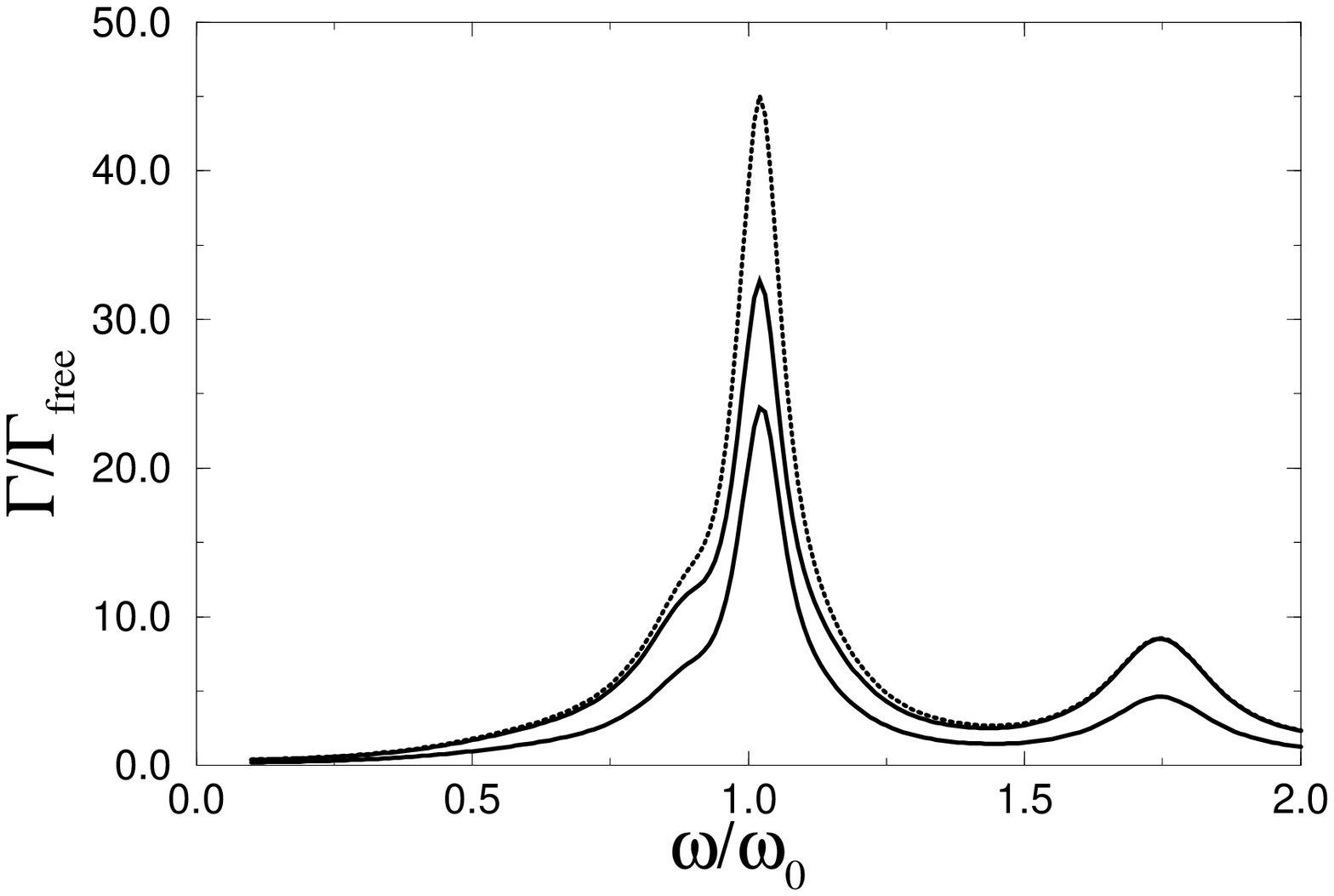}}
\end{center}
\caption{Same as in Fig. 3. but for the Onsager cavity radius 
$R_c=0.03\lambda$.}
\end{figure}

With decreasing the Onsager cavity radius, the absorption contribution
to the decay rate becomes gradually the dominant contribution 
with the leading near-field term equal to the nonradiative 
rate of the molecule. In Fig. 4 we represent the situation 
where $R_c$ is still large enough, so that nonradiative and radiative 
contributions to $\hat{\Gamma}_{\rm loc}$ are of the same order of 
magnitude. In this regime, the decay rate may approximately 
be written as
\begin{equation}
\hat{\Gamma}_{\rm loc}\approx{\cal L}_{\rm Ons}
\left[\frac{\varepsilon''}{|\varepsilon|^2}(k_0R_c)^{-3}+\eta+
\hat{\Gamma}^{\rm sc}\right].
\end{equation}
The difference between $\hat{\Gamma}_{\rm loc}$ (upper solid line) 
and $\hat{\Gamma}$ (lower solid line) this time arises not only 
because of the overall extra factor ${\cal L}_{\rm Ons}$, but also 
because of the $3/2$ times larger nonradiative contribution to 
$\hat{\Gamma}$ [cf. Eqs. (\ref{g0loc}) and (\ref{gama0})]. This 
$3/2$-factor in the corresponding nonradiative rates also leads 
to a significantly larger decay rate $\hat{\tilde{\Gamma}}_{\rm loc}$ 
(dotted line) than the true rate $\hat{\Gamma}_{\rm loc}$ (upper 
solid line). We note, however, that the comparison between 
$\hat{\Gamma}_{\rm loc}$, $\hat{\Gamma}$, and 
$\hat{\tilde{\Gamma}}_{\rm loc}$ is given here only for illustrative 
purposes as their difference is a consequence of our arbitrary 
input: $R_m=R_c$. Indeed, in addition to demanding, on physical 
grounds, the same order of magnitude for $R_m$ and $R_c$, any 
other relation between these parameters may also be assumed. 

Of course, owing to the strong increase of the 
$R_c$-dependent terms, for even smaller $R_c$ the decay rate 
$\Gamma^0_{\rm loc}$ largely exceeds the cavity-induced rate
$\Gamma^{\rm sc}_{\rm loc}$, so that 
$\Gamma_{\rm loc}\approx\Gamma^0_{\rm loc}$ holds. 
For a detailed discussion of the decay rate in this regime, we therefore 
refer the reader to Ref. \cite{scheel2}. 

\section{Summary}
In this work we have calculated the decay rate and the classical
external power loss of an excited molecule located in the center of 
an absorbing and dispersive dielectric sphere by adopting 
the Onsager (real cavity) model for the local field. We have found 
that the external fields of a dipole classically representing the 
molecule calculated with and without the Onsager cavity scale with 
$3\varepsilon(\omega)/[3\varepsilon(\omega)+1]$. This 
immediately gives $|3\varepsilon(\omega)/[3\varepsilon(\omega)+1]|^2$
as the correct local-field correction factor for the (radiation 
and absorption) power loss of the molecule outside the sphere
in the absorbing case. Whereas this result could have been guessed
on the basis of a straightforward analytical continuation of the 
corresponding result in the lossless case, the local-field corrections 
to the total decay rate (power loss) of the molecule are found to be 
much more complex. The total decay rate is found to consists of the 
decay rate for the infinite sphere, as very recently obtained for 
an absorbing medium in Ref. \cite{scheel2}, and of the cavity-induced 
rate, for which we have obtained a similar expression. When expressed
in terms of the Green function for the sphere, these results 
become formally system-independent. This suggests the general results 
for the local-field corrections to the decay rate and to the 
external power loss of a molecule in an absorbing cavity and 
located away from the cavity walls.

\appendix
\section{Dipole power loss}
To calculate $W^0$ using the Poynting's theorem, it is convenient to  
rewrite the dipole electromagnetic field in spherical coordinates.
With $\tensor{\bf I}=\hat{\bf r}\hat{\bf r}+
\hat{\bbox \theta}\hat{\bbox \theta}+
\hat{\bbox \phi}\hat{\bbox \phi}$ in Eq. (\ref{E0}), letting 
the origin at the dipole site and assuming, for simplicity, 
that ${\bf p}=p\hat{\bf z}$, we have for $r\neq 0$ ($k_0=\omega/c$)
\begin{mathletters}
\begin{eqnarray}
{\bf E}^0({\bf r};\omega)&=&ikk_0^2p
\left\{\frac{2h^{(1)}_1(kr)}{kr}\cos\vartheta\hat{\bf r}
\right.\nonumber\\
&+&\left.\left[\frac{h^{(1)}_1(kr)}{kr}
-h^{(1)}_0(kr)\right]\sin\vartheta\hat{\bbox \theta}\right\},
\label{E0sph}
\end{eqnarray}
\begin{equation}
{\bf B}^0({\bf r};\omega)=k^2k_0ph^{(1)}_1(kr)
\sin\vartheta\hat{\bbox \phi},
\label{B0sph}
\end{equation}
\end{mathletters}
where
\[h^{(1)}_0(z)=-i\frac{e^{iz}}{z},\;\;\;
h^{(1)}_1(z)=-\frac{e^{iz}}{z}(1+\frac{i}{z}),\]
are the spherical Hankel functions of the first kind.
The radial component of the Poynting's vector
\[{\bf P}({\bf r};\omega)=\frac{c}{8\pi}{\rm Re}
{\bf E}^0({\bf r};\omega)\times[{\bf B}^0({\bf r};\omega)]^*\]
is then easily found to be
\begin{eqnarray}
\hat{\bf r}\cdot{\bf P}({\bf r};\omega)&=&\frac{c}{8\pi}{\rm Im}
\frac{|k|^6|{\bf p}|^2}{\varepsilon\sqrt{\varepsilon^*}}
\left[-\frac{h^{(1)}_1(kr)}{kr}\right.\nonumber\\
&+&\left.h^{(1)}_0(kr)\right]\left[h^{(1)}_1(kr)\right]^*
\sin^2\vartheta.
\label{Prad}
\end{eqnarray}
This determines the angular distribution of the energy flow
$dW^0_f/d\Omega=r^2\hat{\bf r}\cdot{\bf P}({\bf r};\omega)$
through a spherical surface of radius $r$ around the dipole.
Upon the integration over the angles, we therefore have
\begin{equation}
W^0_f(r)=\frac{\omega|{\bf p}|^2}{3}
\frac{\varepsilon''}{|\varepsilon|^2}
\frac{|(1-ikr)e^{ikr}|^2}{r^3}
+\eta\frac{\omega k_0^3|{\bf p}|^2}{3}e^{-2k''r}.
\label{W0flow}
\end{equation}

The dipole energy absorbed per second within the volume $V_r$ of the
sphere is given by
\begin{equation}
W^0_a(r)=\frac{\omega\varepsilon''}{8\pi}\int_{V_r}d^3{\bf r}
|{\bf E}^0({\bf r};\omega)|^2.
\label{Wabs}
\end{equation}
With
\begin{eqnarray}
|{\bf E}^0({\bf r};\omega)|^2&=&\frac{|k|^6|{\bf p}|^2}
{|\varepsilon|^2}
\left[4\left|\frac{h^{(1)}_1(kr)}{kr}\right|^2
\cos^2\right.\vartheta\nonumber\\
&+&\left.\left|\frac{h^{(1)}_1(kr)}{kr}
-h^{(1)}_0(kr)\right|^2\sin^2\vartheta\right]\nonumber
\end{eqnarray}
and performing the angular integration, we have
\begin{eqnarray}
W^0_a&&(r)=\frac{\omega|{\bf p}|^2}{3}
\frac{\varepsilon''}{|\varepsilon|^2}\int^r_{R_c} dr
\left[\frac{3}{r^4}|(1-ikr)e^{ikr}|^2\right.\nonumber\\
&&-\left.2\frac{k'^2-k''^2-k''|k|^2r}{r^2}e^{-2k''r}+
|k|^4e^{-2k''r}.\right]
\end{eqnarray}
Partially integrating the first term and noticing that the remaining 
integrals involving the dipole near field cancel, we obtain 
\begin{eqnarray}
W^0_a(r)&=&\frac{\omega|{\bf p}|^2}{3}
\frac{\varepsilon''}{|\varepsilon|^2}\left[-\frac{|(1-ikr)e^{ikr}|^2}
{r^3}\right]^r_{R_c}\nonumber\\
&+&\eta\frac{\omega k_0^3|{\bf p}|^2}{3}
(e^{-2k''R_c}-e^{-2k''r}).
\label{W0abs}
\end{eqnarray}
This, together with Eq. (A2), gives Eq. (\ref{Wpoy}) for the total 
dipole power loss $W^0_f(r)+W^0_a(r)$.

\section{Dipole field in multilayered spherical media}

Generalizing the result for an infinite medium [Eq. (\ref{B0sph})], 
the magnetic field of a dipole oscillating along the $z$-axis in the 
center of an $N$-layered spherical system [Fig. 1] can be written 
in the form 
\begin{equation}
{\bf B}({\bf r};\omega)=\varepsilon_1k^3_0pf(r)\sin\vartheta\hat{\bbox \phi}.
\label{Bsph}
\end{equation}
Through 
\[{\bf E}({\bf r};\omega)=\frac{i}{k_0\varepsilon(r)}
\nabla\cdot{\bf B}({\bf r};\omega),\] 
the electric field is therefore given by
\begin{eqnarray}
{\bf E}({\bf r};\omega)&=&ik^2_0p\frac{\varepsilon_1}
{\varepsilon(r)}
\left[\frac{2f(r)}{r}\cos\vartheta\hat{\bf r}\right.\nonumber\\
&-&\left.\frac{[rf(r)]'}{r}\sin\vartheta\hat{\bbox \theta}\right],
\label{Esph}
\end{eqnarray}
with the prime denoting the derivative of the function in the
brackets. The general form of the function $f(r)$ in the $l$th 
layer is 
\begin{equation}
f_l(r)=h^{(1)}_1(k_1r)\delta_{l1}+C^N_{l+}h^{(1)}_1(k_lr)+
C^N_{l-}h^{(2)}_1(k_lr),
\end{equation}
where $h^{(i)}_1(z)$ are spherical Hankel functions. The last 
two terms here are the solutions of the homogeneous Maxwell's 
equations \cite{jackson} and give the scattered field 
${\bf E}^{sc}({\bf r};\omega)$ in the system. The regularity of 
this field at the origin demands that $C^N_{1+}=C^N_{1-}\equiv C^N_1/2$, 
while the outgoing wave condition at infinity demands that 
$C^N_{N-}=0$. The rest of the coefficients $C^N_{l\pm}$ are 
determined from the boundary conditions 
\[f(r_l^-)=f(r_l^+)\;\;\;{\rm and}\;\;\;
\frac{[rf(r)]'_{r_l^-}}{\varepsilon(r_l^-)}=
\frac{[rf(r)]'_{r_l^+}}{\varepsilon(r_l^+)}\]
at the layer interfaces.
The scattered field in the central layer is therefore 
generally given by 
\begin{eqnarray}
{\bf E}_1^{sc}({\bf r};\omega)&=&ik_1k_0^2pC^N_1
\left[\frac{2j_1(k_1r)}{k_1r}\cos\vartheta\hat{\bf r}
\right.\nonumber\\
&-&\left.\frac{[k_1rj_1(k_1r)]'}{k_1r}\sin\vartheta
\hat{\bbox \theta}\right],
\label{E1sc}
\end{eqnarray}
where $j_1(z)$ is the spherical Bessel function. In the other 
layers it is given by (${\bf E}_l={\bf E}_l^{sc}$)
\begin{eqnarray}
{\bf E}_l({\bf r};\omega)&=&ik_1k_0^2p
\sqrt{\frac{\varepsilon_1}{\varepsilon_l}}
\left\{C^N_{l+}\left[\frac{2h^{(1)}_1(k_lr)}{k_lr}
\cos\vartheta\hat{\bf r}\right.\right.\nonumber\\
&-&\left.\frac{[k_lrh^{(1)}_1(k_lr)]'}{k_lr}
\sin\vartheta\hat{\bbox \theta}\right]\nonumber\\
&&+C^N_{l-}\left[\frac{2h^{(2)}_1(k_lr)}{k_lr}
\cos\vartheta\hat{\bf r}\right.\nonumber\\
&-&\left.\left.\frac{[k_lrh^{(2)}_1(k_lr)]'}{k_lr}
\sin\vartheta\hat{\bbox \theta}\right]\right\},
\label{El}
\end{eqnarray}
with $C^N_{N-}=0$. Specially, since $j_1(z)/z\rightarrow 1/3$ and 
$[zj_1(z)]'/z \rightarrow 2/3$ for small $z$, the scattered field in the 
center of the system is given by
\begin{equation}
\left.{\bf E}_1^{sc}({\bf r};\omega)\right|_{r\rightarrow 0}
=ik_1k_0^2C^N_1\frac{2}{3}{\bf p}.
\label{E1sc0}
\end{equation}

For $N=2$, we find ($\rho_i=k_ir_1$)
\begin{eqnarray}
C^2_1(\varepsilon_1,\varepsilon_2;r_1)&=&
\frac{1}{D}\left[\varepsilon_2h^{(1)}_1(\rho_2)
[\rho_1h^{(1)}_1(\rho_1)]'\right.\nonumber\\
&-&\left.\varepsilon_1h^{(1)}_1(\rho_1)
[\rho_2h^{(1)}_1(\rho_2)]'\right],
\label{C12}
\end{eqnarray}
and
\begin{equation}
C^2_{2+}(\varepsilon_1,\varepsilon_2;r_1)=\frac{i\varepsilon_2}
{\rho_1D},
\label{C22}
\end{equation}
where
\begin{equation}
D=\varepsilon_1j_1(\rho_1)[\rho_2h^{(1)}_1(\rho_2)]'-
\varepsilon_2h^{(1)}_1(\rho_2)[\rho_1j_1(\rho_1)]'.
\end{equation}

For $N=3$, we have ($\rho_{ij}=k_ir_j$)
\begin{eqnarray}
&&C^3_1(\varepsilon_1,\varepsilon_2,\varepsilon_3;r_1,r_2)=
\nonumber\\
&&\frac{1}{j_1(\rho_{11})}\left[\frac{\beta_2h^{(1)}_1(\rho_{21})-
\beta_1h^{(2)}_1(\rho_{21})}
{\alpha_1\beta_2-\alpha_2\beta_1}-h^{(1)}_1(\rho_{11})\right],
\label{C13}
\end{eqnarray}

\begin{mathletters}
\begin{eqnarray}
C^3_{2+}(\varepsilon_1,\varepsilon_2,\varepsilon_3;r_1,r_2)&=&
\frac{\beta_2}{\alpha_1\beta_2-\alpha_2\beta_1},\\
C^3_{2-}(\varepsilon_1,\varepsilon_2,\varepsilon_3;r_1,r_2)&=&
\frac{-\beta_1}{\alpha_1\beta_2-\alpha_2\beta_1},
\end{eqnarray}
\end{mathletters}
\begin{equation}
C^3_{3+}(\varepsilon_1,\varepsilon_2,\varepsilon_3;r_1,r_2)=
-\frac{i\varepsilon_3}{\rho_{22}}\frac{2}
{\alpha_1\beta_2-\alpha_2\beta_1},
\label{C33}
\end{equation}
where
\begin{eqnarray}
\alpha_j&=&-\frac{i\rho_{11}}{\varepsilon_2}
[\varepsilon_1j_1(\rho_{11})[\rho_{21}h^{(j)}_1(\rho_{21})]'
\nonumber\\&-&
\varepsilon_2h^{(j)}_1(\rho_{21})[\rho_{11}j_1(\rho_{11})]'],
\label{alphaj}
\end{eqnarray}
and
\begin{eqnarray}
\beta_j&=&\varepsilon_3h^{(1)}_1(\rho_{32})
[\rho_{22}h^{(j)}_1(\rho_{22})]'\nonumber\\
&-&\varepsilon_2h^{(j)}_1(\rho_{22})
[\rho_{32}h^{(1)}_1(\rho_{32})]'.
\label{betaj}
\end{eqnarray}

\end{document}